%% file: S5v2.tex
\documentclass[12pt]{article}

\include{def}
\usepackage{amsmath}
\def\ds{\displaystyle}

\def\CH{\widetilde{\cal{H}}}
\def\HCH{\widehat{\cal{H}}}
\def\c{\cal{C}}

\def\btheta{\bar{\theta}}
\def\CF{\cal{F}}

\def\wp{\widetilde{P}}

\def\CV{\cal{V}}
\def\CY{\cal{Y}}
\def\CZ{\cal{Z}}
\def\a{\underline{a}}
\def\b{\underline{b}}
\def\c{\underline{c}}

\begin{document}
\vspace*{-.6in}
\thispagestyle{empty}
\begin{flushright}
USB-preprint: SB/F/299-02
\end{flushright}
\baselineskip = 20pt

\vspace{.5in}
{\Large
\begin{center}
{Super Five Brane Hamiltonian and the Chiral Degrees of Freedom}
\end{center}}

\begin{center}
A. De Castro{\footnote{E-mail address:
adecastr@pion.ivic.ve}$^{\dagger}$} and A.
Restuccia{\footnote{E-mail address: arestu@usb.ve}$^{\ddag}$}\\
\emph{$^{\dagger}$Instituto Venezolano de Investigaciones
Cient\'{\i}ficas
\\ {\bf(IVIC)} , Centro de F\'{\i}sica, AP 21827, Caracas 1020-A, Venezuela.\\
{$^{\ddag}$ Universidad Sim\'on Bol\'{\i}var, Departamento de F\'{\i}sica,
\\ AP 89000, Caracas 1080-A, Venezuela}}
\end{center}
\vspace{.5in}

\begin{center}
\textbf{Abstract}
\end{center}

\begin{quotation}
\noindent We construct the Hamiltonian of the super five brane in
terms of its physical degrees of freedom. It does not depend on
the inverse of the induced metric. Consequently, some singular
configurations are physically admissible, implying an
interpretation of the theory as a multiparticle one.  The
symmetries of the theory are analyzed from the canonical point of
view in terms of the first and second class constraints. In
particular it is shown how the chiral sector may be canonically
reduced to its physical degrees of freedom.
\end{quotation}
\vfil

\newpage

\pagenumbering{arabic}

\section{\label{intro}Introduction}

One of the hopeful models to understand the origin of superstring
 dualities is the conjectured M-theory in eleven dimensions,
 where there are only two extended objects
allowed by  supersymmetry: the super 2-brane (supermembrane) and
the super 5-brane. The supermembrane theory has  been widely
studied during the last years, see for example:
\cite{Nicolai:1998ic},  \cite{deWit:1987iw}, \cite{deWit:1988ig},
\cite{deWit:1989ct}, \cite{hopetesis}, \cite{Bergshoeff:1988in},
\cite{Bergshoeff:1987dh}, \cite{Bergshoeff:1987cm},
\cite{Duff:1988cs}, \cite{deWit:1990vb},
\cite{GarciadelMoral:2001zb}, \cite{Boulton:2001gz}. Nevertheless,
the analysis of the covariant and hamiltonian quantum dynamics of
the super 5-brane is now in its dawn, \cite{Pasti2}, \cite{JS3},
\cite{townsendund} \cite{lamamiaqui}, \cite{DeCastro:2001gp},
between others. In particular, we would like to understand the
nature of the super five brane spectrum. In 1997, a manifestly
covariant action for the super 5-brane was constructed  in
\cite{Pasti2}. Independently, at the same time, a non manifestly
covariant action was obtained by J. Schwarz et. al. \cite{JS3}.
The field equations were first obtained in \cite{Howe:1997yn} and
analyzed in \cite{Howe:1997fb}, \cite{Bergshoeff:1998vx}. More
recently, we analyzed some dynamical aspects for the M5-brane
`bosonic sector' \cite{lamamiaqui}\cite{DeCastro:2001gp}, it
included a complete study of the canonical structure of the
bosonic sector of the M5-brane starting from the PST action in the
gauge where the scalar field is fixed as the world volume time. We
found a quadratic dependence on the antisymmetric field for the
canonical Hamiltonian. This formulation contains second class
constraints that we removed preserving the locality of the field
theory in order to construct a master action with first class
constraints only. The  algebra of the 6 dimensional
diffeomorphisms generated by the first class constraints was
explicitly obtained.  We constructed the nilpotent BRST charge of
the theory and its BRST invariant effective theory. Finally, we
obtained its physical Hamiltonian and analyzed its stability
properties.

In this work we extend the analysis to the super 5-brane theory.
In particular we show that the canonical lagrangian of the super
5-brane may be formulated without the assumption of the existence
of the inverse of the induced metric, which is a requirement of
the original PST as well as the Schwarz  et.al actions.
Consequently the hamiltonian formulation of our theory admits as
physical configurations ones where locally the determinant of the
induced metric in zero. They allow to connect disjoint 5-branes by
singular 1,2,3 and 4-branes without changing the energy of the
original configurations . They lead then to the interpretation of
super 5-brane as a multiparticle theory, in a similar way as the
existence of the string-like spikes suggest the same
interpretation for the supermembrane \cite{deWit:1989ct},
\cite{Nicolai:1998ic}. The canonical study of the  PST super
5-brane action, in the gauge in which the  auxiliar scalar field
is equal to the world volume time, shows a mixture of first and
second class constraints which includes the expected
reparametrization and kappa symmetry generators and the second
class fermionic constraint, besides the first and second class
constraints associated to the antisymmetric field. The canonical
Hamiltonian is quadratic in the antisymmetric gauge field. It is
very interesting to observe that the mixture of first and second
class constraints associated to the  chiral field gauge symmetry
may be decoupled from the rest of the constraints. This feature
allow us to remove the second class constraints of this sector and
construct a master canonical action. From this supersymmetric
formulation, we can recover the bosonic master formulation found
in \cite{DeCastro:2001gp}. It is important to comment that the
Hamiltonian for the super 5-brane was first obtained in
\cite{Bergshoeff:1998vx} . It apparently depends on the
determinant of $(g+\CH)$, which depends on the fourth power of
$\CH$, however, a further combinations of the terms in their
constraints yields exactly the same constraints we  obtain in this
paper. It was raised in \cite{Bergshoeff:1998vx} the problem of
the presence of the second class constraints in the chiral gauge
field sector. In this paper we explicitly revolves that problem
with the construction of our master Hamiltonian. In the last
section we find the light cone gauge Hamiltonian for the theory
and analyze its stability properties. The canonical analysis of
the theory requires the explicit form of all terms in the
Lagrangian, therefore, the explicit form of the Wess-Zumino term
of the PST \cite{Pasti2} super five brane Lagrangian is obtained
as a first step.

\section{\label{super}The Super 5-Brane Action}
The super 5-brane PST Lagrangian is given by
\begin{equation}
L=L_1 + L_2 + L_{3},
\end{equation}
with:
\begin{equation}
\begin{split}
L_1&=2\sqrt{-\det M_{MN}}\; d\sigma^{0}\wedge\cdots\wedge
d\sigma^{5}\\ L_2&={1\over 2(\partial a)^2} {\CH}^{MN}
{\cal{H}}_{MNP} G^{PL}
\partial_L a\;d\sigma^{0}\wedge\cdots\wedge d\sigma^{5}\\ L_3&=
\Omega_6.
\end{split}
\end{equation}
where,
\begin{equation*}
\begin{split}
M_{MN}&= G_{MN} + i {G_{MP} G_{NL}\over\sqrt{-G (\partial a)^2}}
\tilde{\cal H}^{PL},\\
{\CH}^{PL}& = {1\over 6} \epsilon^{PLQMNR} {\cal{H}}_{MNR}
\partial_Q a.\\
{\cal{H}}_{MNP}&=H_{MNP}-b_{MNP}
\end{split}
\end{equation*}
are the  supersymmetric extensions for the Born-Infeld type term
and the antisymmetric field strength $H=dB$ respectively. In
components, $H$ and $B$ may be written as:
\begin{equation*}\label{}
\begin{split}
H&={1\over{3!}}
H_{MNL}d\sigma^{M}\wedge d\sigma^{N}\wedge d\sigma^{L}\\
&={1\over{3!}}(\partial_{M}B_{NL}+\partial_{L}B_{MN}+\partial_{N}B_{LM})
d\sigma^{M}\wedge d\sigma^{N}\wedge d\sigma^{L}\\
B&={1\over{2!}}B_{MN}d\sigma^{M}\wedge d\sigma^{N}
\end{split}
\end{equation*}
and
\begin{equation*}\label{b}
b=\frac{1}{6}\btheta\Gamma_{ab}d\theta[dX^ a dX^ b +\Pi^a dX^b
+\Pi^a \Pi^b].
\end{equation*}
where $\Pi^a$ is the SUSY-invariant Cartan form which is given by:
\begin{equation*}
\Pi^{a}=\Pi^{a}_{M}d\sigma^{M}=(\partial_M
X^{a}+\bar{\theta}\Gamma^{a}\partial_M\theta)d\sigma^M
\end{equation*}
and the induced supermetric by:
\begin{equation*}
G_{MN}=\Pi^ a_M\Pi^ b_N\eta_{ab}
\end{equation*}
$a,b=0,\cdots,10$ are the Minkowski  space-time indices while
$M,N=0,\cdots,5$ are the world volume indices. The Wess--Zumino
term $\Omega_6 $ is determined by the closed seven-form $I_7 = d
\Omega_6$, where
\begin{equation*}
\begin{split}
I_7&= - {1\over 2} {\cal H}\wedge d \bar\theta\Gamma_{ab}
d\theta\wedge\Pi^a\wedge\Pi^b\\ &+ {1\over 60} d \bar\theta
\Gamma_{abcdef}d\theta\wedge\Pi^a\wedge\Pi^b\wedge\Pi^c\wedge\Pi^d\wedge\Pi^e\wedge\Pi^f\nonumber.
\label{I7cov}
\end{split}
\end{equation*}
$I_7$ and  $\Omega_6$ may be expressed without the use of the
induced metric $G_{MN}$ nor the scalar field $a$. Moreover, the
explicit expression of $\Omega_6$ was not needed in order to prove
the global supersymmetry and the local $\kappa$-symmetry of the
action. In our analysis however, which will be based in the
construction of the physical hamiltonian of the super 5-brane, it
is required to have an explicit expression for the $\Omega_6$
term. It is given, up to a closed six-form, by:
\begin{equation*}
\begin{split}
\Omega_6&=dB\wedge b-{1\over 60}\btheta\Gamma_{abcde}d\theta
dX^adX^bdX^cdX^ddX^e\\&-{1\over
24}d\btheta\Gamma_{abcde}\theta\wedge d\btheta\Gamma^e\theta\wedge
dX^adX^bdX^cdX^d\\ &+ {1\over 12} \btheta\Gamma_{ab}d\theta
d\btheta\Gamma_{cd}\theta\wedge d\btheta\Gamma^d\theta\wedge
dX^adX^bdX^c
\\&-{1\over 18} d\btheta\Gamma_{abcde}\theta\wedge
d\btheta\Gamma^e\theta\wedge d\btheta\Gamma^d\theta\wedge
dX^adX^bdX^c\\ &- {1\over 24}\btheta\Gamma_{cd}d\theta\wedge
dX^cdX^d \btheta\Gamma_{ab}d\theta\wedge
\btheta\Gamma^ad\theta\wedge \btheta\Gamma^bd\theta \\&+{1\over
24}d\btheta\Gamma_{abcde}\theta\wedge d\btheta\Gamma^e\theta\wedge
d\btheta\Gamma^d\theta\wedge d\btheta\Gamma^c\theta\wedge
dX^adX^b\\ &+{1\over 60}d\btheta\Gamma^b\theta\
d\btheta\Gamma^a\theta \wedge d\btheta\Gamma_{ab}\theta \wedge
d\btheta\Gamma_{cd}\theta \wedge d\btheta\Gamma^c\theta \wedge
dX^d\\&+{1\over 60}d\btheta\Gamma_{abcde}\theta\wedge dX^a\wedge
\btheta\Gamma^bd\theta\wedge \btheta\Gamma^cd\theta \wedge
\btheta\Gamma^dd\theta\wedge \btheta\Gamma^ed\theta\\&-{1\over
360}d\btheta\Gamma_{abcde}\theta\wedge
d\btheta\Gamma^{e}\theta\wedge d\btheta\Gamma^{d}\theta\wedge
d\btheta\Gamma^{c}\theta\wedge d\btheta\Gamma^{b}\theta\wedge
d\btheta\Gamma^{a}\theta.
\end{split}
\end{equation*}

\section{Canonical Analysis and the Master Supersymmetric Hamiltonian}
In this section we consider the construction of the canonical
formulation of the PST theory, in the gauge in which the auxiliar
scalar field $a$ is equal to the world volume time and the time
components of the antisymmetric field $B_{0\mu}$ are zero
($\mu=1,\cdots, 5$ denote the spatial world volume indices), and
its hamiltonian. After fixing the light cone gauge and the gauge
symmetries related to the antisymmetric field, we will obtain the
physical hamiltonian of the theory. We will show that is possible
to perform a canonical reduction to the light cone gauge, as is
usual in string and supermembrane theory. Moreover, we will show
that the gauge symmetry related to the antisymmetric field may
also be fixed in a way which allows a canonical reduction to the
physical degrees of freedom. In fact, in order to do so, we
realize that the selfduality condition on the curvature of the
antisymmetric field in six dimensions describes a first order
propagating equation for six physical degrees of freedom. It is
then natural to look for a canonical formulation where those
degrees of freedom should be realized in terms of canonically
conjugate fields.

We will consider the supersymmetric extension of an $ADM$
parametrization of the metric \cite{ADM61} as in
\cite{DeCastro:2001gp} to obtain:
\begin{eqnarray}
L&=&2n\sqrt{M} -\frac{1}{4}N^\rho
\widehat{\cal{V}}_\rho+\frac{1}{2}\widetilde{H}^{\mu\nu}\partial_0{B}_{\mu\nu}\\
&&-{\CH^{\mu\nu}}b_{\mu\nu0}+{\CF}(X,\theta)\nonumber
\end{eqnarray}
where
\begin{eqnarray*}
g&=&\det{g_{\mu\nu}}, \;\; g_{\mu\nu}=G_{\mu\nu}\cr\cr
M&=&1+\widehat{\CY}+\widehat{\CZ}\cr\cr \widehat{\CY}&=&{1\over
2}g^{-1}{\CH^{\mu\nu}}{\CH_{\mu\nu}}\cr\cr \widehat{\CZ}&=&{1\over
64}g^{-1}g^{\mu\nu}\widehat{\CV}_\mu\widehat{\CV}_\nu
\end{eqnarray*}
and
\[
\widehat{\CV}_\mu=\epsilon_{\mu\alpha\beta\gamma\delta}{\CH^{\alpha\beta}}{\CH^{\gamma\delta}}
\]
The spatial world volume indices are raised and lowered with the
induced metric $g_{\mu\nu}$. It will turn out that the final
expression of the hamiltonian does not require the existence of
the inverse  of $g_{\mu\nu}$. ${\CF}(X,\theta)$ in the lagrangian
density denotes the contribution from $\Omega_6-{1\over
2}\epsilon^{\mu\nu\alpha\beta\rho}b_{\mu\nu\alpha}b_{\beta\rho0}$
 independent on the antisymmetric field. The latest term is the remaining part of
 $L_2$. It involve at most
linear terms on the time derivative of the $X^a$ and $\theta$,
since it is a six form constructed from $dX$ and $d\theta$. The
conjugate momenta to $X^a$ may be directly evaluated.It is:
\begin{equation*}\label{}
P_a=\wp_a+f_a(X^a,\theta)
\end{equation*}
where
\begin{equation*} \wp_a ={2\over
n}\sqrt{gM}\left[-{\dot{\Pi}}_a+N^{\alpha}\Pi_{a\alpha}\right]-{1\over
4}\widehat{\CV}^\rho\Pi_{a\rho},
\end{equation*}
and
\begin{equation*}
f_a=\ds\frac{\delta{\CF}}{\delta \dot{X^a}},
\end{equation*}
from which we deduce the following constraints:
\begin{eqnarray}
\widehat{\Phi}
&=&{1\over 2}\wp_a\wp^a+2g(1+{\widehat{\CY}})=0\label{phi}\\
\widehat{\Phi}_\alpha
&=&\wp_a\Pi^a_\alpha+\frac{1}{4}{\widetilde{\CV}}_\alpha=0\label{phia}
\end{eqnarray}
The conjugate momenta to the $B_{\mu\nu}$ will be denoted
$P^{\mu\nu}$ and satisfies the constraint
\begin{equation}\label{antisym}
\Omega^{\mu\nu}\equiv P^{\mu\nu}-\widetilde{H}^{\mu\nu}=0
\end{equation}
which is exactly the same constraint that appears in the bosonic
5-brane canonical formalism. The previous constraints (\ref{phi})
and (\ref{phia}) are modified with respect to the bosonic ones by
the terms $f_a$ and $b_{\mu\nu\lambda}$, while (\ref{antisym})
remains unchanged. This property of the antisymmetric field has
important consequences in the construction of the physical
hamiltonian.  At this point, we would like to comment that we have
obtained in \cite{congreso} a preliminar canonical version with
\[P^{\mu\nu}-{\CH}^{\mu\nu}=0
\] that arises from the fact that the first term in $\Omega_6$ may be
written as: $-B\wedge db$. Both theories are consistent,
nevertheless, the present version of this constraint is more
manageable for the later analysis of the chiral degrees of
freedom. This behavior can be attributed to the existence of a
canonical transformation between both versions. Finally, the
fermionic constraint arises directly from the evaluation of the
conjugate momenta $\xi$ to $\theta$. It is:
\begin{equation}\label{fermionic}
\Psi\equiv\xi-(\btheta\Gamma^a \wp_a-{\CH^{\mu\nu}}\frac{\delta
b_{\mu\nu0}}{\delta\dot{\theta}}+\frac{\delta
{\CF}}{\delta\dot{\theta}})=0
\end{equation}

(\ref{phi}), (\ref{phia}), (\ref{antisym}) and (\ref{fermionic})
are the complete set of constraints of the super 5-brane theory.
The canonical hamiltonian is a linear combination of these
constraints,
\begin{equation}\label{canham}
\rm{\bf{H}}=\Lambda\widehat{\Phi}+\Lambda^\mu\widehat{\Phi}_\mu+\Lambda_{\mu\nu}\Omega^{\mu\nu}+\Psi\eta
\end{equation}
The set of constraints is a mixture of second and first class one,
as usual for the superstring and supermembrane theories.

We may now introduce a new canonical lagrangian with a gauge group
which contains as a subgroup the one generated by the first class
constraints of the previous formulation (\ref{canham}), such that
under partial gauge fixing it reduces to (\ref{canham}). To do so,
we first replace ${\CH}^{\mu\nu}$ in (\ref{phi}), (\ref{phia}) and
(\ref{fermionic}) by:
\begin{equation}\label{}
{\CH}^{\mu\nu}\longrightarrow\widehat{\cal{H}}^{\mu\nu}\equiv{1\over
2}(P^{\mu\nu}+\widetilde{H}^{\mu\nu})-\widetilde{b}^{\mu\nu}
\end{equation}
which is a valid procedure under (\ref{antisym}). Having done that
replacement we now relax (\ref{antisym}) into:
\begin{eqnarray}
\Omega^\nu&\equiv&\partial_\mu P^{\mu\nu}=0\label{FC1}\\
\Omega^{5i}&\equiv &P^{5i}-\widetilde{H}^{5i}=0\quad
i=1,\cdots,4\label{FC2}
\end{eqnarray}
and consider them together with:
\begin{eqnarray} {\Phi}
&=&{1\over 2}\wp_a\wp^a+2g(1+{\CY})=0\label{newphi}\\
{\Phi}_\alpha
&=&\wp_a\Pi^a_\alpha+\frac{1}{4}{\CV}_\alpha=0\label{newphia}\\
\Psi &\equiv &\xi-(\btheta\Gamma^a
\wp_a-{\HCH^{\mu\nu}}\frac{\delta
b_{\mu\nu0}}{\delta\dot{\theta}}+\frac{\delta
{\CF}}{\delta\dot{\theta}})=0\label{newfermionic}
\end{eqnarray}
Here ${\CV}_\mu$, $\CY$ and $\CZ$ now depend on ${\HCH^{\mu\nu}}$
instead of ${\CH}^{\mu\nu}$. We notice that (\ref{FC1}) and
(\ref{FC2}) commute between themselves and with: (\ref{newphi}),
(\ref{newphia}) and (\ref{newfermionic}). They are then, first
class constraints. Moreover under partial gauge fixing of the
gauge symmetry they generate we may recover (\ref{antisym}), and
consequently the hamiltonian (\ref{canham}). We thus conclude
that:
\begin{equation}\label{canhamFC}
\rm{\bf{H}}=\Lambda{\Phi}+\Lambda^\mu{\Phi}_\mu+\rho_{\mu}\Omega^{\mu}+\rho_{5i}\Omega^{5i}+\eta\Psi
\end{equation}
defines a master canonical system describing the super 5-brane
theory, where the constraints related to the antisymmetric field
have been raised to first class ones. If we turn  the fermionic
coordinates off in (\ref{canhamFC}), it  reduces to the master
bosonic hamiltonian obtained in \cite{DeCastro:2001gp}. We will
show in the next section that we can use the additional gauge
symmetry generated by (\ref{FC2}) to perform a canonical reduction
of the system ending up with a canonical description of the super
5-brane action in terms of the physical degrees of freedom of the
antisymmetric field.

\section{The Physical Hamiltonian}
The constraints (\ref{newphi}), (\ref{newphia}) and
(\ref{newfermionic})   are a mixture of first and second class
constraints. The first class ones generates the six dimensional
difeomorphisms on the world volume, together with the
$\kappa$-symmetry. We consider now the light cone gauge fixing
conditions:
\begin{eqnarray}
X^+&=&-P^+_0\tau,\label{lcg1}\\ P^+&=&\sqrt{\omega}P^+_0,\label{lcg2}\\
\Gamma^+\theta &=&0\label{ksymfix}
\end{eqnarray}
where $\omega$ is a time independent scalar density on the world
volume. In general, it is not possible to impose the condition
$\omega=1$ on the whole world volume, we prefer then to leave it
explicitly in the expression. The physical consequences of the
theory should be independent of $w$, which may be thought as the
determinant of a metric on the world volume. In this sense, the
dependence on $\omega$ is like the dependence on the metric in
topological field theories, the metric appears through the gauge
fixing procedure but the observables of the theory are independent
of it.

The gauge fixing (\ref{lcg1}), (\ref{lcg2}),  (\ref{ksymfix})
together with the constraints (\ref{newphi}), (\ref{newphia}) and
(\ref{newfermionic}) allow a canonical reduction of the
hamiltonian (\ref{canhamFC}). In fact, the canonical conjugate
pairs:
\begin{equation*}
\begin{split}
X^+,&\quad P_+\\ X^-,&\quad P_-\\
\Gamma^+\theta,&\quad \xi\Gamma^-
\end{split}\end{equation*}
may be eliminated from the above mentioned gauge fixing conditions
and constraints. One is left only with the constraints:
\begin{equation}\label{C1}
\Psi\Gamma^+=\xi\Gamma^+ -(\btheta\Gamma^I
P_I-{\HCH^{\mu\nu}}\frac{\delta
b_{\mu\nu0}}{\delta\dot{\theta}}+\frac{\delta
{\CF}}{\delta\dot{\theta}})\Gamma^+ =0
\end{equation}
and
\begin{equation}\label{C2}
\Theta_{\mu\nu}\equiv\partial_{[\mu}\left(\displaystyle\frac{\widetilde{P}_I\partial_{\nu]}\Pi^I}{\omega}+{1\over
4
\omega}{\CV}_{\nu]}-P_0^+\btheta\Gamma^-\partial_{\nu]}\theta\right)=0
\end{equation}
where $I,J= 1,2,\cdots,9$.

A suitable linear combination of the left handed sides of
(\ref{C1}) and (\ref{C2}) defines the volume preserving
diffeomorphisms generator, the fermionic constraint
$\Psi\Gamma^+=0$ is left as a second class one.

The light cone gauge Hamiltonian then reads:
\begin{equation}\begin{split}
{{\bf{H}}}_{LCG} &=\ds\frac{1}{\sqrt{\omega}}[{1\over
2}\wp^J\wp_J+2g(1+{\cal{Y}})]
+P^+_0f^-(X^a,\theta)\\&+\rho_{\mu}\Omega^{\mu}+\rho_{5i}\Omega^{5i}+\Lambda^{\alpha\beta}
\Theta_{\alpha\beta}+{\Psi\Gamma^+}\eta
\end{split}\label{LCG}
\end{equation}
We may now consider a further canonical reduction related to
(\ref{FC2}). We impose the gauge fixing condition:
\begin{equation}\label{GFonB1}
B_{5i}=0
\end{equation}
that allows to eliminate the canonical conjugate pair
$(B_{5i},P^{5i})$ using the first class constraint (\ref{FC2}).
The remaining constraint (\ref{FC1}) reduces then to:
\begin{equation}\label{}
\partial_iP^{ij}+\partial_5\widetilde{H}^{5j}=0
\end{equation}
which may be rewritten as:
\begin{equation}\label{FC3}
\partial_j(P^{ij}-\widetilde{H}^{ij})=0
\end{equation}
In order to disentangle the physical degrees of freedom the
antisymmetric field we may impose a further gauge fixing condition
associated to the first class constraint (\ref{FC3}). We consider:
\begin{equation}\label{GFonB2}
B_{4{\a}}=0
\end{equation}
where ${\a}=1,2,3$. This is an admissible gauge fixing condition
which has the interesting property that the kinetic term
$P^{ij}\dot{B}_{ij}$ reduces completely. To do so, we notice that
(\ref{FC3}) may be resolved explicitly:

\begin{equation}\label{}
P^{ij}=\widetilde{H}^{ij}+\epsilon^{ijkl}\partial_kA_l
\end{equation}
where $A_l$ are the remaining degrees of freedom of the momenta
$P^{ij}$. If we now evaluate the kinetic term we obtain:
\begin{equation*}\label{}
\begin{split}
\langle P^{ij}\dot{B}_{ij} \rangle &=\langle
\epsilon^{ijkl}\partial_5B_{kl}\dot{B}_{ij}+\dot{A}_l
\epsilon^{ijkl}\partial_kB_{ij} \rangle\\ &=\langle \dot{A}_l
\epsilon^{ijkl}\partial_kB_{ij} \rangle
\end{split}
\end{equation*}
Since (\ref{GFonB1}) and (\ref{GFonB2}) ensures that the first
term on the right hand side member is zero. We then define:
\begin{equation*}\label{}
\begin{split}
P^{\a}&\equiv -\epsilon^{\a\b\c}B_{\b\c}\\
P&\equiv\epsilon^{\a\b\c}\partial_{\a}B_{\b\c}\\
\end{split}
\end{equation*}
they satisfy the equation:
\begin{equation}\label{remanente}
\partial_{\a}P^{\a}+P=0
\end{equation}
It yields:
\begin{equation*}\label{}
\langle P^{ij}
\dot{B}_{ij}\rangle=\langle\dot{A}_{\a}\partial_4P^{\a}+\dot{A}_4P
\rangle
\end{equation*}
which implies that $\partial_4P^{\a}$ and $P$ are the conjugate
momenta to $A_{\a}$ and $A_4$ respectively. We may perform a final
canonical reduction by imposing the gauge fixing $A_4=0$ and
eliminate its conjugate momentum $P$ from (\ref{remanente}). All
the dependence of the hamiltonian on the antisymmetric field is
through the terms:
\begin{equation*}\label{}
\begin{split}
{\HCH}^{ij}&\equiv {1\over 2} (P^{ij}+\widetilde{H}^{ij}) +\tilde{b}^{ij}\\
{\HCH}^{5i}&\equiv{\widetilde{\cal{H}}}^{5i}
\end{split}
\end{equation*}
which appear quadratically in the hamiltonian.

We notice that:
\begin{equation*}\label{}
\begin{split}
{\CH}^{\a\b}&=0\\
{\CH}^{4\a}&=-{1\over 6}\epsilon^{\a\b\c}\partial_5B_{\b\c}={1\over 6}\partial_5P^{\a}\\
{\CH}^{54}&=-{1\over 6}\epsilon^{\a\b\c}\partial_{\a}B_{\b\c}={1\over 6}\partial_{\a}P^{\a}\\
{\CH}^{5\a}&={1\over
6}\epsilon^{\a\b\c}\partial_4B_{\b\c}=-{1\over6}\partial_4P^{\a},
\end{split}
\end{equation*}
hence the canonical lagrangian can be expressed in terms of
$A_{\a}$ and  $P^{\a}$  which describe the unrestricted
independent degrees of freedom associated to the antisymmetric
field. The explicit terms in the canonical lagrangian are:
\begin{equation}\label{}
\begin{split}
&\langle\dot{A}_{\a}\partial_4
P^{\a}+{\HCH}^{ij}{\HCH}^{kl}g_{ik}g_{jl}+\\
&+4{\HCH}^{i5}{\HCH}^{kl}g_{ik}g_{5l} +
4{\HCH}^{i5}{\HCH}^{j5}(g_{ij}g_{55}-g_{i5}g_{j5})\rangle
\end{split}
\end{equation}
where
\begin{equation*}\label{}
\begin{split}
{\HCH}^{\a\b}&={1\over 2}\epsilon^{\a\b kl}\partial_kA_{l}+\tilde{\b}^{\a\b}\\
{\HCH}^{4\a}&={1\over 12}\partial_5P^{\a}-{1\over 2}\epsilon^{\a\b\c}\partial_{\b}A_{c}+\tilde{\b}^{4\a}=\\
{\HCH}^{54}&={1\over 6}\partial_{\a}P^{\a}+\tilde{\b}^{54}\\
{\HCH}^{5\a}&=-{1\over6}\partial_4P^{\a},
\end{split}
\end{equation*}
\section{Discussion and Conclusions}

We constructed the physical hamiltonian of the super 5-brane
theory which is still constrained by the second class fermionic
constraint and by the first class volume preserving
diffeomorphisms. We prefer to leave this gauge symmetry without
fixing it, since it may have a relevant  interpretation in terms
of a noncommutative geometry. That was the case for the
supermembrane. In that case the area preserving diffeomorphisms on
the spatial world volume is the same as the symplectomorphisms
which are very closely related to the noncommutative geometry
constructed in terms of the Weyl  algebra bundle
\cite{Martin:2001zv}. In our case, the group of symplectomorphisms
is contained in  the volume preserving diffeomorphisms. A complete
analysis of that relation seems to be very important.

The canonical lagrangian we have obtained is expressed in terms of
the covariant induced metric $g_{\mu\nu}$, we do not require the
existence of the inverse contravariant metric in our construction
of the hamiltonian of the super 5-brane theory. The original PST
formulation as well as the Schwarz et. al. construction requires
the uses of the inverse metric $g^{\mu\nu}$ in order to define
their lagrangians. This property which was already discuss for the
bosonic sector of the 5-brane in \cite{DeCastro:2001gp}, and shown
to be valid even in the realization of the algebra of
diffeomorphisms  in terms of the first class constraints of the
theory, remains valid for the supersymmetric formulation of the
5-brane. It has the important consequence that singular
configurations which annihilate the determinant of the induced
metric at a neighborhood of any point on the world volume are
admissible configurations of the theory. They are essential in the
physical interpretation of the supermembrane as a multiparticle
theory \cite{deWit:1988ig} \cite{Nicolai:1998ic}. In that case
they are string like spikes which can change the topology of any
configuration and connect  disjoint membranes without changing the
energy of the system. Consequently, these configurations are
responsible, together with the supersymmetry of the continuos
spectrum of the supermembrane. In the super 5-brane case those
singular configurations  may be $1,2,3,\mbox{or},4$-branes which
also provide the same interpretation of the theory as a
multiparticle one. This property was analyzed in
\cite{DeCastro:2001gp} and we have shown now that the same
analysis may be extended to the supersymmetric 5-brane theory.

In order to describe the self duality property of the super
5-brane equations of motion in a canonical local formulation we
propose the hamiltonian (\ref{LCG}) together with the constraints
(\ref{C1}), (\ref{C2}) and (\ref{FC3}). This formulation may be
further reduced preserving the canonical structure in terms of the
conjugate pairs $(A_a, \partial_4P^a)$, $(A_4, P)$, however, this
latest formulation becomes non local because of the term
$\partial_5P^a=\partial_5\partial_4^{-1}(\partial_4P^a)$. In this
case we can performed a final canonical reduction by imposing the
gauge fixing condition  $A_4=0$ and eliminating its conjugate
momentum $P$ from (\ref{LCG}) and end up with the unconstrained
pair $(A_{\a}, \partial_4P^{\a})$.

If we consider the dimensional reduction of the super 5-brane
hamiltonian identifying  $X^5=\sigma^5$, and taking all other
$\partial_5\cdot=0$ we end up with the super $4$-brane hamiltonian
in 10 dimensions Minkowski space. In that case the nonlocal term
of the  canonical formulation becomes zero and we obtain a local
canonical lagrangian for the super 4-brane. The terms involving
the antisymmetric field becomes now quadratic, because $P_5$ has
to be eliminated from the constraints:
\[P_5+{1\over 4}{\CV}_5=0\]
\[
\]
hence the hamiltonian incudes the term:
\[ {1\over 16} {\CV}_5^2 \]
where
\[
{\CV}_5=\epsilon_{5ijkl}\left({{P^{ij}}\over2}+\tilde{b}^{ij}
\right)\left({{P^{kl}}\over2}+\tilde{b}^{kl} \right)
\]

\newpage 
\bibliography{S5v1bib}


\appendix
\include{apend}

\end{document}

%% file: apend.tex
\section*{\label{Omega}APPENDIX A: Majorana spinors and Fierz identities}

{Majorana spinors:}
\begin{equation*}
\begin{split} \btheta\Gamma_a\chi&=-\bar{\chi}\Gamma_a\theta\\
\btheta\Gamma_{ab}\chi&=-\bar{\chi}\Gamma_{ab}\theta\\
\btheta\Gamma_{abcde}\chi&=-\bar{\chi}\Gamma_{abcde}\theta
\end{split}
\end{equation*}
{Fierz identities:}
\begin{equation*}
\begin{split}
d\btheta\Gamma^a d\theta d\btheta\Gamma_{ab}+
d\btheta\Gamma_{ab}d\theta d\btheta\Gamma^a&=0\\ d\btheta\Gamma^e
d\theta d\btheta\Gamma_{abcde}+ d\btheta\Gamma_{abcde}d\theta
d\btheta\Gamma^e&=6\btheta\Gamma_{[ab}d\theta d\btheta\Gamma_{cd]}
\end{split}
\end{equation*}

The following identities can be deduced from Fierz identities:
\begin{eqnarray*}
d(d\btheta\Gamma^a\theta
d\theta\Gamma_{ab}\theta)&=&2d\btheta\Gamma^a d\theta\wedge
d\btheta\Gamma_{ab}\theta\cr d(d\btheta\Gamma^a\theta\wedge
d\btheta\Gamma^b\theta\wedge d\btheta\Gamma_{ab}\theta)&=&
-3d\btheta\Gamma^a\theta\wedge d\btheta\Gamma^b\theta\wedge
d\btheta\Gamma_{ab}d\theta
\end{eqnarray*}
\begin{equation*}
d(d\btheta\Gamma_{abcde}\theta\wedge d\btheta\Gamma^e\theta)
=-2(d\btheta\Gamma_{abcde}d\theta)
d\btheta\Gamma^e\theta+6\btheta\Gamma_{[ab}d\theta
d\btheta\Gamma_{cd]}\theta
\end{equation*}
\begin{eqnarray*}
d(d\btheta\Gamma_{abcde}\theta \wedge d\btheta\Gamma^e\theta\wedge
d\btheta\Gamma^e\theta\wedge d\btheta\Gamma^d\theta )&=&
-3d\btheta\Gamma_{abcde}d\theta\wedge d\btheta\Gamma^e\theta\wedge
d\btheta\Gamma^d\theta\cr &&+ 12\btheta\Gamma_{[ab}d\theta
d\btheta\Gamma_{cd]}\theta\wedge d\btheta\Gamma^d\theta
\end{eqnarray*}
\begin{eqnarray*}
d(d\btheta\Gamma_{abcde}\theta \wedge d\btheta\Gamma^e\theta\wedge
d\btheta\Gamma^e\theta\wedge d\btheta\Gamma^d\theta \wedge
d\btheta\Gamma^c\theta)&=& -4d\btheta\Gamma_{abcde}d\theta\wedge
d\btheta\Gamma^e\theta\wedge d\btheta\Gamma^d\theta \wedge
d\btheta\Gamma^c\theta\cr &&+ 18\btheta\Gamma_{[ab}d\theta
d\btheta\Gamma_{cd]}\theta\wedge d\btheta\Gamma^d\theta\wedge
d\btheta\Gamma^c\theta
\end{eqnarray*}
\begin{equation*}
\begin{split}
d(d\btheta\Gamma_{abcde}\theta \wedge d\btheta\Gamma^e\theta\wedge
d\btheta\Gamma^e\theta\wedge d\btheta\Gamma^d\theta \wedge &
d\btheta\Gamma^c\theta\wedge d\btheta\Gamma^b\theta)=\\&=
-5(d\btheta\Gamma_{abcde}d\theta) d\btheta\Gamma^e\theta\wedge
d\btheta\Gamma^d\theta \wedge d\btheta\Gamma^c\theta\wedge
d\btheta\Gamma^b\theta\\ &+ 24(\btheta\Gamma_{[ab}d\theta
d\btheta\Gamma_{cd]}\theta) d\btheta\Gamma^d\theta\wedge
d\btheta\Gamma^c\theta\wedge d\btheta\Gamma^b\theta
\end{split}
\end{equation*}
\begin{equation*}
\begin{split}
d(d\btheta\Gamma_{abcde}\theta \wedge d\btheta\Gamma^e\theta\wedge
 d\btheta\Gamma^d\theta \wedge
d\btheta\Gamma^c\theta\wedge & d\btheta\Gamma^b\theta\wedge
d\btheta\Gamma^a\theta)=\\&=-6(d\btheta\Gamma_{abcde}d\theta)
d\btheta\Gamma^e\theta\wedge d\btheta\Gamma^d\theta \wedge
d\btheta\Gamma^c\theta\wedge d\btheta\Gamma^b\theta \\&+
30(\btheta\Gamma_{[ab}d\theta d\btheta\Gamma_{cd]}\theta)
d\btheta\Gamma^d\theta\wedge d\btheta\Gamma^c\theta\wedge
d\btheta\Gamma^b\theta\wedge d\btheta\Gamma^a\theta
\end{split}
\end{equation*}